\documentclass[10pt,journal,compsoc]{IEEEtran}

\usepackage{amsmath, amssymb}
\usepackage{graphicx}
\usepackage{booktabs}
\usepackage[hidelinks]{hyperref}
\usepackage{subcaption}
\usepackage{xcolor}
\usepackage{tcolorbox}
\usepackage{float}
\usepackage{dsfont}
\usepackage{tabularx}
\usepackage{booktabs}
\usepackage{array}

\usepackage[nocompress]{cite}

\title{Multi-Scale Temporal Structure of the Conti Ransomware Communication Network}

\author{Cameron~Cornell, Matthew~Roughan, Lewis~Mitchell,
Chad~Whelan, David~Bright, and Callum~Jones%
\IEEEcompsocitemizethanks{%
\IEEEcompsocthanksitem C. Cornell, M. Roughan, and L. Mitchell
are with the Adelaide Data Science Centre, School of Mathematical Sciences,
Adelaide University, SA 5005, Australia.
\IEEEcompsocthanksitem C. Whelan, D. Bright, and C. Jones
are with the School of Humanities and Social Sciences,
Deakin University, Australia.
\IEEEcompsocthanksitem This study was funded by the Australian Government
through the Office of National Intelligence and Security Discovery Research
Grant (NISDRG) program [NI2301000042].
}}

\begin{document}


\IEEEtitleabstractindextext{%
\begin{abstract}

Communication networks provide a means of inferring broader organisational structure from accumulated local interactions.
Cybercriminal organisations are particularly vulnerable to such inference, despite their secretive nature, because their distributed operation necessitates coordination through centralised digital communication infrastructure. The public leak of the Conti ransomware organisation's internal communications provides a rare longitudinal record of this coordination.
Yet common approaches to analysing such longitudinal records aggregate interactions into a static network or model the communication process as stationary, both assuming a temporal homogeneity rarely tested for criminal organisations.
Here we show substantial temporal heterogeneity at each scale examined in the Conti record, and demonstrate that much of it can be accounted for through appropriate, diagnostically guided choices of event representation and observation period.
Collapsing rapid message bursts into conversational episodes brings communication timing closer to a constant-intensity reference and clarifies longer-timescale structure, which resolves into a sequence of distinct communication regimes, including a major transition following a coordinated platform migration. Even so, within these regimes heterogeneity persists, most evident in churn among peripheral participants, while highly active actors remain consistently observable.
Together, these findings motivate explicit consideration of multi-scale temporal dynamics in the analysis, modelling and longitudinal monitoring of cybercriminal networks.

\end{abstract}


}

\maketitle
\IEEEdisplaynontitleabstractindextext



\IEEEraisesectionheading{
    \section{Introduction}\label{sec:introduction}
}

Networks of communication appear across many technological and sociological systems and provide a natural framework for studying interactions between communicating entities ~\cite{wasserman1994,newman2018networks}. Examples range from telecommunications infrastructure ~\cite{onnela2007structure,faloutsos1999power} and online social platforms ~\cite{leskovec2008planetary,kwak2010twitter} to organisational and collaborative systems ~\cite{diesner2005enron,kossinets2006empirical}. While individual communications primarily facilitate local information exchange and coordination, the aggregate patterns formed by many such interactions often reveal higher-level properties of the underlying system. Organisational structure, influential actors, and patterns of collective behaviour may emerge from the accumulation of many individual interactions, despite not being explicitly encoded within any single message. Communication is therefore not merely a by-product of organisation, but one of its primary observable manifestations, providing a window into coordination, hierarchy, and collective behaviour.

Cybercriminal organisations are a particularly interesting example of communication systems \cite{Sparrow1991TheAO,Morselli2009Inside}. Unlike many traditional criminal networks, where relationships must often be inferred indirectly from arrests, transactions, co-offending records, or surveillance data, cybercriminal activity is comparatively centralised through digital communication platforms. The geographically distributed nature of modern ransomware operations necessitates sustained remote coordination between specialised actors, making communication infrastructure a primary mechanism through which organisational activity occurs. While communication records do not necessarily capture all interactions between participants, they provide a comparatively rich vantage on the interaction process. Consequently, communication data offer a valuable opportunity to directly examine the structure and dynamics of cybercriminal organisations through the interactions that underpin their operation.

A unique opportunity to study such communication processes emerged in early 2022 following the public leak of internal communications from the Conti ransomware organisation \cite{gray2022money,paternoster2025inside,whelan2025internal}. Conti was among the largest and most active ransomware groups operating during this period, with activities spanning ransomware deployment, extortion, and supporting cybercriminal services. Following public statements made by the group during the Russia--Ukraine conflict, a large archive of internal communications was released online by an apparent insider.
The resulting dataset contains 168,740 timestamped messages exchanged among 450 distinct pseudonymous user IDs (hereafter referred to as actors) over a twenty-month period. This combination of network size, temporal depth, and direct communication records provides an unusually detailed view of the internal operation of a large cybercriminal organisation.




Communication records are often analysed in an aggregated manner, providing a static view of the underlying organisational topology. 
Yet the interactions generating these networks can change substantially even during routine operation: actors enter and leave the system, participation levels fluctuate, and interaction structures reconfigure through time. Consequently, the observed communication network may be more appropriately viewed as a dynamically reconfiguring system than a single static structure, motivating the need for temporally resolved analyses of organisational behaviour \cite{holme2012temporal,masuda2016guide}. As in communication network analysis more broadly, existing analyses of cybercriminal communication networks have been conducted predominantly on temporally aggregated representations, limiting the extent to which these evolving communication processes can be directly observed. 


To address this limitation, we investigate temporal heterogeneity within the Conti communication network through multiple lenses corresponding to different resolutions of temporal activity.

\begin{itemize}

\item \textbf{Conversational microstructure.} We examine burst communication dynamics and show how conversational activity can produce substantial temporal clustering at the message level. We introduce a simple burst-collapse procedure that merges rapid within-dyad exchanges into conversational episodes, and show that this representation is more closely aligned with the underlying interaction process, allowing genuine network dynamics to be distinguished from microscale conversational effects. This clarity is apparent in the subsequent regime analysis, where burst collapse increases mean within-regime adjacency correlation by approximately 24\%. Comparing the collapsed process with a Poisson null model across thresholds further provides a simple criterion for choosing the event timescale on which subsequent analyses are based.


\item \textbf{Communication regimes.}
We identify four distinct communication regimes from changes in communication structure over time. These regimes differ in their interaction patterns and levels of organisational activity. Viewing the network as a collection of 14-day windows, we show that the identified regimes decompose a global mean cross-window correlation of 0.211 into 0.428 within regimes and 0.085 between them, revealing periods of comparatively stable communication structure. We further examine the relationship between regime transitions and observed operational events, including evidence linking a major structural transition to a coordinated platform migration.




\item \textbf{Participation dynamics and observability.} We examine actor lifespans, turnover, and discovery to assess how organisational participation is observed through time. Highly active actors are discovered rapidly and consistently, whereas lower-activity participants continue to be discovered throughout each regime at a rate that cannot be explained by their lower aggregate activity alone. Thus, a limited observation period may capture the principal communicators while substantially underrepresenting the network periphery.
\end{itemize}


Collectively, these analyses provide a temporally resolved view of cybercriminal communication, highlighting the need to account for changes in communication structure, participant activity, and organisational coordination through time.




\section{Related Work}

\subsection{Network Analysis of Criminal Organisations}

The application of network analysis to criminal organisations has a long history within criminology, originating with early assessments of its prospects for criminal intelligence \cite{Sparrow1991TheAO}, and developing into a substantial body of work characterising the structure of criminal groups, identifying strategically positioned actors, and informing disruption strategies \cite{Morselli2009Inside, bright2012}. A persistent theme within this literature concerns the observational limitations inherent to criminal network data, which must typically be reconstructed from sources such as arrest records, surveillance data, or wiretaps, yielding incomplete representations with uncertain boundaries \cite{Sparrow1991TheAO, Campana2012}. Communication records provide comparatively direct observation, capturing the interaction process itself rather than requiring relationships to be inferred from co-occurrence or association \cite{Campana2012}. The majority of such research, however, examines criminal networks at a single point in time. 
A smaller longitudinal literature has documented structural change in drug trafficking networks across multi-year periods \cite{bright2013, Morselli01052007} and modelled network formation using stochastic actor-oriented approaches \cite{bright2019}, demonstrating that criminal networks reorganise substantially in response to operational demands and law enforcement pressure \cite{Morselli01052007, Brightgreenhill}. The temporal resolution available in such studies nevertheless remains coarse, with networks aggregated into a small number of multi-month or multi-year waves dictated by data availability rather than by the underlying dynamics of the organisation. Assumptions of statistical stationarity are thus often made before any detailed mathematical modelling.

The growth of cybercrime has extended this research programme to online criminal networks, where digital trace data offer substantially richer observation. Network analytic methods have been applied to hacker communities \cite{decaryhet},  darknet markets \cite{duxbury2018}, and website defacement ecosystems \cite{perkins2023}, while organisational research has characterised profit-driven cybercriminal groups as increasingly firm-like enterprises exhibiting internal division of labour \cite{lusthaus}. Even where longitudinal digital records are available, however, analyses have generally proceeded through comparisons of discrete network snapshots rather than temporally resolved treatment of the underlying interaction stream \cite{perkins2023}.

\subsection{Analyses of the Conti Leak}

\begin{table*}[t]
\centering
\caption{Comparison of Conti studies and their treatment of temporal structure.}
\label{tab:conti_literature_comparison}
\small
\renewcommand{\arraystretch}{1.2}

\begin{tabularx}{\textwidth}{
    >{\raggedright\arraybackslash}p{0.15\textwidth}
    >{\raggedright\arraybackslash}p{0.25\textwidth}
    >{\raggedright\arraybackslash}X
}

\toprule
\textbf{Study} &
\textbf{Emphasis} &
\textbf{Temporal considerations} \\
\midrule

Earlier Conti studies
\cite{gray2022money,paternoster2025inside,whelan2025internal}
&
Characterise Conti's operations, hierarchy and organisational roles.
&
Primarily aggregate network and content analysis, with
conversational preprocessing in \cite{paternoster2025inside}.
\\

\addlinespace

Manatova et al. (2026)
\cite{manatova2026organizational}
&
Analyse organisational structure and resilience.
&
Group messages into conversations to weight an aggregate
undirected network, with segmentation evaluated through
manual assessment.
\\

\addlinespace

Sadewo et al. (2026)
\cite{sadewo2026managing}
&
Explain changes in managerial communication using
relational event models.
&
Merge message fragments less than one minute apart and
compare managerial effects across roughly equal-length
periods selected for model stability.
\\

\addlinespace

\textbf{Present study}
&
Systematically characterise temporal heterogeneity and
its consequences for network analysis.
&
Assess how conversational aggregation changes interarrival
statistics and subsequent analyses; identify communication
regimes and relate transitions to organisational events;
examine remaining temporal heterogeneity through participant
discovery within regimes.
\\

\bottomrule
\end{tabularx}
\end{table*}

The 2022 leak of internal Conti communications has generated a small but growing body of research examining the internal operation of a large ransomware enterprise. Existing studies have used the leaked corpus to trace ransom payments and characterise the group's business model \cite{gray2022money}, examine money laundering strategies \cite{nazzari2023payday}, identify discussion topics through machine learning \cite{ruellan2024conti}, and map the group's organisational structure through combined network and content analysis \cite{paternoster2025inside, whelan2025internal}. This latter work reveals a hierarchical, business-like organisation with defined leadership roles and task specialisation, mirroring structures observed in legitimate enterprises \cite{paternoster2025inside}.

These studies establish the Conti corpus as a uniquely detailed observational record of cybercriminal organisation. Earlier analyses have, however, been predominantly conducted on temporally aggregated representations, often projecting the full twenty-month communication record onto a single static structure. Given the evidence from the broader criminal networks literature that such organisations reconfigure substantially through time \cite{bright2013, Morselli01052007}, aggregate representations risk conflating distinct organisational phases within a single summary. Recent studies increasingly incorporate temporal considerations \cite{manatova2026organizational,sadewo2026managing}, often in support of applied organisational analysis, highlighting the need for a systematic investigation of the effects of temporal heterogeneity in these data. The present study systematically examines these choices: how conversational aggregation changes the observed temporal statistics, how comparatively stable communication regimes can be identified, and what temporal heterogeneity remains within them. Table~\ref{tab:conti_literature_comparison} summarises the emphasis and temporal considerations of these studies.





\subsection{Temporal Networks and Human Communication Dynamics}

Beyond criminology, a substantial methodological literature has developed around temporal networks, in which the timing of interactions is retained as an explicit element of the representation rather than projected out through aggregation \cite{holme2012temporal,masuda2016guide}. A central finding of this literature is that the choice of temporal aggregation is consequential, with the structural properties of aggregated networks depending systematically on the length and placement of the aggregation window \cite{krings2012effects, sulo2010meaningful}, and no single natural aggregation timescale generally exists, as relevant dynamics occur across multiple temporal resolutions \cite{holme2012temporal, krings2012effects}. Related work has further shown that finite observation windows systematically distort estimated interaction statistics, including inter-event time distributions \cite{kivela2015estimating}. These considerations directly motivate the multi-resolution approach adopted in this study, in which conversational microstructure, regime-level reorganisation, and participation dynamics are treated as distinct phenomena operating at distinct timescales.

At the shortest timescales, human communication has been established to be temporally inhomogeneous. Inter-event time distributions in human activity exhibit heavy tails inconsistent with memoryless constant-rate behaviour, a phenomenon attributed variously to priority-queueing mechanisms in task execution \cite{barabasi2005origin} and to the superposition of periodic activity cycles with cascades of correlated activity \cite{malmgren2008poissonian}. The cascading process perspective of \cite{malmgren2008poissonian}, in which a nonhomogeneous Poisson process governs the initiation of active intervals within which multiple events occur in rapid succession, is particularly relevant to the present work. The burst-collapse procedure developed in \autoref{sec:burst_collapse} operationalises a closely related two-timescale view nonparametrically, treating rapid successive messages as conversational episodes whose initiation, rather than internal reply structure, constitutes the organisationally meaningful event. Where parametric temporal structure is required, self-exciting point process models provide the natural formal counterpart \cite{hawkes1971point, laub2021elements}, though such models are not pursued here.


A natural null model against which we assess these dynamics is a dyadic Poisson model, in which each dyad is assigned a fixed baseline interaction intensity and events occur independently at constant rate. This provides a constant-rate benchmark whose deviations isolate the temporal structure of interest at both conversational and organisational scales. Identifying the boundaries of the communication regimes examined in \autoref{sec:regimes} relates more broadly to change-point detection in networks, for which a developing methodological literature exists \cite{peel2015detecting}, while the participant discovery analyses of \autoref{sec:accum} draw on species richness estimation techniques originating in ecology \cite{good1953population, chao1984nonparametric, colwell1994estimating} and applied within criminology to the estimation of hidden populations \cite{macdonald2017shuffle}.
\section{Data}

\label{sec:data}

The dataset comprises a leaked corpus of internal communications from the Conti cybercriminal organisation over the period 21 June 2020 to 2 March 2022, publicly available as English-translated Jabber chat CSVs from Northwave Security \cite{northwave_conti_chats}. It contains 168,740 timestamped, directed messages among 450 distinct pseudonymous user IDs. For each message, the dataset records a source and target identifier, a timestamp, and both encoded and partially decoded message content. We treat each user ID as a network node and refer to these IDs as actors throughout, although an individual may use more than one ID.


The communications were conducted via an internal Jabber/XMPP-based messaging infrastructure, commonly used in cybercriminal operations due to its support for private, self-hosted servers. Messages are stored as logs extracted from this system, with timestamps typically recorded at sub-minute resolution. The leak is widely understood to have originated from internal dissent within the group, resulting in the public release of chat logs and associated operational data. Use of these publicly released Conti data was approved by Deakin University's higher-than-low-risk Human Research Ethics Committee (approval number 2023-165) on the grounds that obtaining consent from members of the Conti chat leaks is impractical and that the benefits of the project outweigh any potential risks. Analyses use pseudonymous user IDs, and no attempt was made to identify individuals. One limitation of the available dataset is a period of discontinuity between 16 November 2020 and 29 January 2021, during which no communication records are available.






Let $V$ denote the set of actors and let
\[
E = \{(i,j,t)\},
\]
for $ {i,j}\in V$, denote the set of timestamped directed communication events, where each element corresponds to a single observed message sent from actor $i$ to actor $j$ at time $t$. Timestamps are re-indexed such that $t=0$ corresponds to the start of the observation period, and are expressed in units of days.

The communication system may therefore be viewed as a temporal directed multigraph, with multiple timestamped edges permitted between any ordered pair of actors. Ordered actor pairs $(i,j)$ define directed dyads, which form the basis for subsequent temporal and interaction-level analysis.


\section{Aggregate Network Structure}


We begin by considering atemporal projections of the communication system, aggregating timestamped message events over specified observation windows. This yields directed network representations in which edge weights correspond to the total number of messages sent between actors during the window.

Formally, for any time window $[t_0,t_1]$, define the weighted adjacency matrix $W^{[t_0,t_1]}$ by
\[
W^{[t_0,t_1]}_{ij}
=
\sum_{(i,j,t)\in E}
\mathds{1}(t_0 \leq t \leq t_1),
\]
so that $W^{[t_0,t_1]}_{ij}$ records the total number of messages sent from actor $i$ to actor $j$ during the interval.

We additionally define the corresponding binary adjacency matrix
\[
A^{[t_0,t_1]}_{ij}
=
\mathds{1}\big(W^{[t_0,t_1]}_{ij} > 0\big),
\]
which records only whether communication occurred, irrespective of message count.

Unless otherwise stated, un-indexed matrices $W$ and $A$ denote aggregation over the full observation period. Summary statistics of the resulting aggregate network are presented in Table~\ref{tab:network_summary}.



\begin{table}[h]
\centering
\caption{Summary statistics of the aggregated communication network.}
\label{tab:network_summary}
\begin{tabular}{l c}
\hline
Statistic & Value \\
\hline
Number of nodes $|V|$ & 450 \\
Number of messages $|E|$ & 168,740 \\
Messages per node $|E|/|V|$ & 375 \\
Average degree & 8.8\\
Median out-degree & 2\\
Median in-degree & 3\\
Active dyads ($W_{ij} > 0$) & 1.96\% \\
Reciprocity & 62.3\% \\
\hline
\end{tabular}
\end{table}

\autoref{tab:network_summary} summarises the aggregated communication network. We define degree in the binary sense, with out-degree and in-degree corresponding to the number of distinct recipients and senders associated with each actor, respectively, irrespective of message volume. In this context, degree reflects the number of unique individuals with whom an actor is observed to communicate at least once. The resulting network is sparse and skewed, with an average degree of 8.8 across 450 nodes, and median out- and in-degrees of 2 and 3, indicating that most actors interact with only a small subset of the network. This is also reflected globally, with only 1.96\% of potential dyads being active.

Reciprocity is defined as the proportion of active dyads exhibiting mutual communication, i.e. the fraction of pairs for which both $W_{ij} > 0$ and $W_{ji} > 0$, conditional on at least one direction being active. The observed reciprocity of 62.3\% indicates that a majority of active connections are bidirectional. This is consistent with the use of the network for coordinated communication, and increases to 72\% when restricted to pairs with more than one observed message, and to  89\% for pairs with more than five messages, indicating that non-reciprocal interactions are concentrated among low-frequency exchanges.




\begin{figure*}[h]
    \centering

    \begin{subfigure}[t]{0.48\linewidth}
        \centering
        \includegraphics[width=\linewidth]{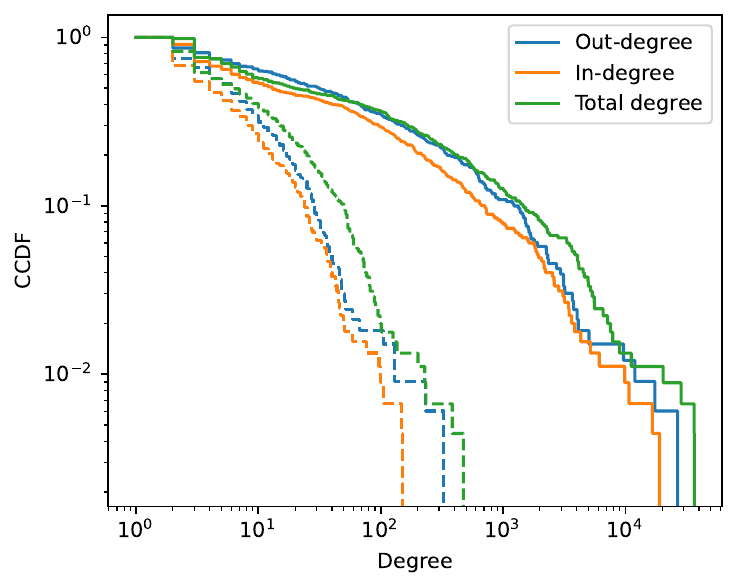}
        \caption{CCDFs of weighted (solid) and binary (dashed) node degree (out-, in-, and total).}
        \label{fig:degree_ccdf}
    \end{subfigure}
    \hfill
    \begin{subfigure}[t]{0.48\linewidth}
        \centering
        \includegraphics[width=\linewidth]{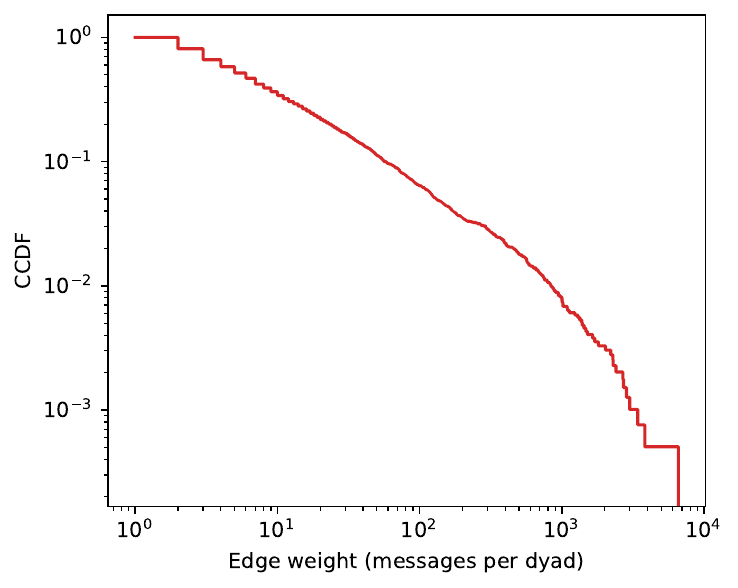}
        \caption{CCDF of edge weight (messages per dyad).}
        \label{fig:edge_weight_ccdf}
    \end{subfigure}

    \caption{CCDFs of node degree and edge weight on log--log scales.}
    \label{fig:basic_ccdf}
\end{figure*}

\autoref{fig:degree_ccdf} displays the complementary cumulative distribution functions (CCDFs) of both binary (dashed) and weighted (solid) out-, in-, and total degree for the communication network. The weighted degree distributions exhibit clear heavy-tailed behaviour, with slow probability decay, though with noticeable curvature relative to the linear form expected under a pure power-law. The corresponding binary degree distributions display a similar overall form, but are concentrated over a substantially smaller range of support. Across both representations, out-, in-, and total degree align closely over the full range, with minimal separation, consistent with the largely reciprocal nature of messaging interactions observed across the network.



In contrast, the edge weight CCDF shown in \autoref{fig:edge_weight_ccdf} exhibits an approximately linear trend over a substantial portion of the support on the log–log scale, consistent with power-law behaviour. This trend truncates in the upper tail, plausibly reflecting a practical limit on sustained communication frequency in a human-driven network. Summary statistics of the distribution indicate substantial heterogeneity, with a mean of 42 messages per active dyad, a median of 5, and a maximum of 6601, consistent with a heavily skewed interaction structure.



\subsection{Poisson Null Model}
\label{sec:poisson_null} 

The aggregate network records the total communication activity associated with each dyad, but does not describe how this activity is distributed through time. To examine the temporal structure of communication separately from differences in interaction frequency, we introduce a Poisson arrival null model.

Under this model, communication between each ordered pair of actors $(i,j)$ is treated as an independent homogeneous Poisson process with dyad-specific rate parameter $\lambda_{ij}$. For a fully observed interval $[t_0,t_1]$, the corresponding message count therefore satisfies
\begin{equation}
    W^{[t_0,t_1]}_{ij}
    \sim \mathrm{Poisson}\bigl(\lambda_{ij}(t_1-t_0)\bigr).
    \label{eq:poisson_null}
\end{equation}
Counts are independent across distinct dyads and across non-overlapping intervals. Communication frequencies may therefore differ substantially across the network, while remaining constant through time within each dyad.

Let $T$ denote the total duration for which communication records are available, excluding the discontinuity described in \autoref{sec:data}. The dyad-specific rates are estimated from the aggregate message counts as
\begin{equation}
    \hat{\lambda}_{ij} = \frac{W_{ij}}{T}.
\end{equation}
The fitted model consequently reproduces the observed aggregate edge weights in expectation, while removing temporal clustering, changes in interaction intensity, and systematic changes in participant activity.

Although clearly simplistic, this model provides a common baseline for the analyses that follow. At short timescales, it implies exponentially distributed interarrival times. At longer timescales, equal-duration observation windows have the same distribution of adjacency structure. Conditional on the aggregate dyad counts, it also provides a baseline for participant discovery obtained by randomly ordering the observed communication events.



\section{Conversational Microstructure}
\label{sec:burst_collapse}

\subsection{Burst Communication}


\begin{figure*}[h]
    \centering

    \begin{subfigure}[t]{0.48\linewidth}
        \centering
        \includegraphics[width=\linewidth]{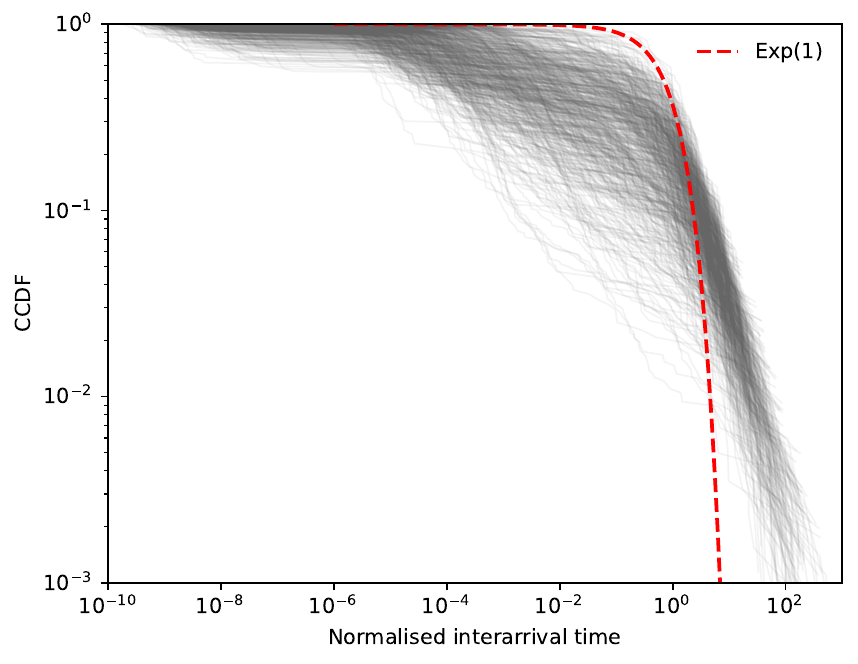}
        \caption{Raw messages.}
        \label{fig:interrivals_raw}
    \end{subfigure}
    \hfill
    \begin{subfigure}[t]{0.48\linewidth}
        \centering
        \includegraphics[width=\linewidth]{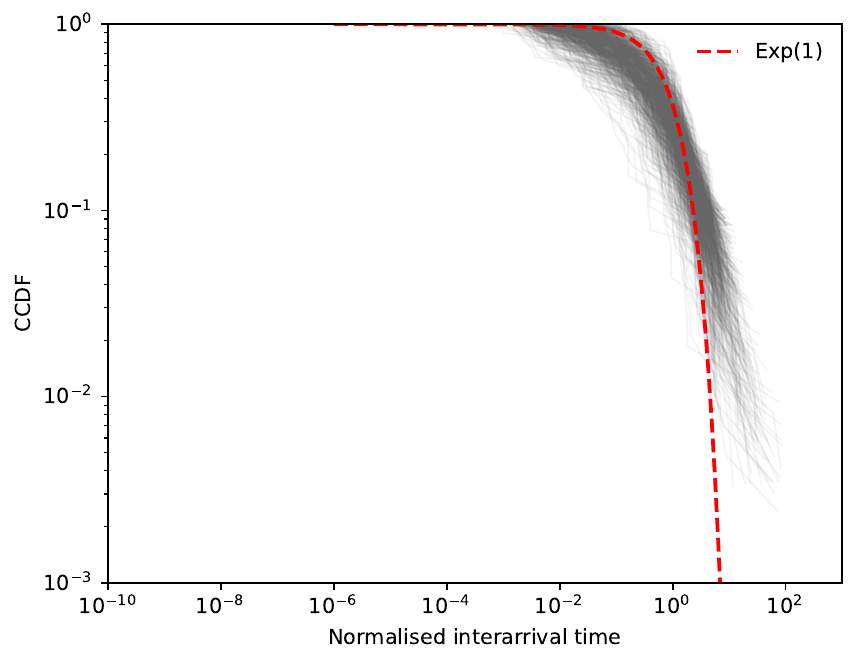}
        \caption{One-hour conversational episodes.}
        \label{fig:interrivals_collapsed}
    \end{subfigure}

    \caption{Dyadic interarrival CCDFs before and after one-hour burst collapse. Interarrivals are normalised by their mean within each directed dyad. Grey curves show the same 672 dyads in both panels; red dashed curves show the Exp(1) reference. Each included dyad has at least 10 interarrivals after collapse.}
    \label{fig:interrivals}
\end{figure*}


\begin{figure*}[h]
    \centering
    \includegraphics[width=0.9\linewidth]{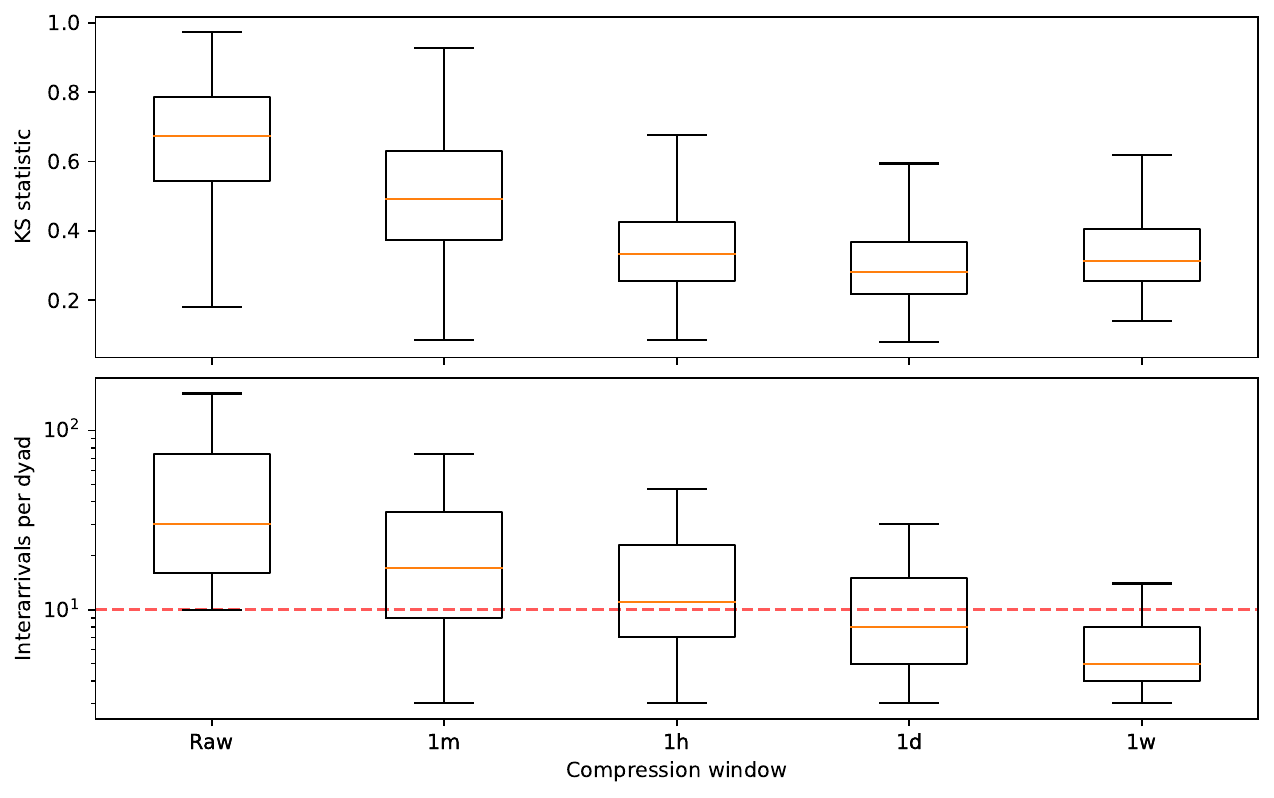}
    \caption{KS statistic relative to Exp(1) and interarrivals per dyad (log scale) under increasing compression thresholds (Raw $\rightarrow$ 1m $\rightarrow$ 1h $\rightarrow$ 1d $\rightarrow$ 1w). The red dashed line marks 10 interarrivals per dyad as a reference.}
    \label{fig:compression}
\end{figure*}

To investigate short-timescale communication dynamics, we compare observed dyadic message timing with the Poisson arrival null model introduced in section~\ref{sec:poisson_null}. Under this model, messages occur independently at a constant dyad-specific rate, implying memoryless communication with no burstiness or short-term dependence beyond the mean interaction frequency. Comparing observed interarrival times with this baseline therefore allows temporal clustering to be examined separately from differences in average communication activity across dyads.





To compare observed communication against this null model, we examine the interarrival times between successive messages within each active dyad. For each ordered dyad $(i,j)$, let
\[
T_{ij} = (t_1,\dots,t_{n_{ij}})
\]
denote the ordered sequence of the $n_{ij}$ message timestamps associated with that dyad. We then define the corresponding interarrival sequence as the successive differences:
\[
\tau_{ij,k} = t_{k+1} - t_k,
\qquad k=1,\dots,n_{ij}-1.
\]

Under the Poisson null model, these interarrival times are exponentially distributed with dyad-specific rate parameter $\lambda_{ij}$,
\[
\tau_{ij,k} \sim \mathrm{Exp}(\lambda_{ij}).
\]

As communication frequencies vary substantially across dyads, direct comparison of raw interarrival times is uninformative. To isolate distributional shape from baseline activity level, we normalise interarrivals within each dyad by their empirical mean,

\begin{equation}
    \tilde{\tau}_{ij,k}
    = \frac{\tau_{ij,k}}{\bar{\tau}_{ij}}.
    \label{eq:normalised_interarrival}
\end{equation}


This rescales each dyad to unit empirical mean interarrival time. 
For a homogeneous Poisson process, rescaling by the true rate gives

\begin{equation}
    \lambda_{ij}\tau_{ij,k}=\frac{\tau_{ij,k}}{E[\tau_{ij,k}]}
    \sim \mathrm{Exp}(1).
\end{equation}

The $\mathrm{Exp}(1)$ distribution therefore provides a common reference for comparing dyads with heterogeneous communication frequencies. Since the mean is estimated from the same interarrivals being compared, the empirical normalisation only approximates this reference, with finite-sample effects becoming more important for dyads with few observations.


\autoref{fig:interrivals_raw} displays the empirical CCDFs of the normalised interarrival times defined in \autoref{eq:normalised_interarrival} across dyads, together with the exponential reference distribution implied by the Poisson null model. Relative to this benchmark, observed interarrival distributions exhibit both substantially greater mass at short interarrival durations and markedly slower tail decay, indicating an excess of both rapid successive messaging and extended inactivity periods compared to the homogeneous Poisson baseline.





\subsection{Burst Collapse}

A wide range of parametric models exist for temporally clustered event sequences, including self-exciting and non-homogeneous process models  \cite{hawkes1971point,laub2021elements} capable of directly capturing burst communication dynamics. However, before turning to such sophisticated temporal models, we first consider a simpler and more intuitive hypothesis: that individual messages may not constitute the appropriate atomic unit of communication.

Specifically, we propose that human communication operates through two distinct temporal regimes: rapid within-conversation dynamics and slower between-conversation initiation dynamics. 
Under this view, multiple messages exchanged in close succession may form part of a single conversational episode. This is common in messaging and email systems, where a single interaction episode may involve multiple messages exchanged in rapid succession before a prolonged period of inactivity. In such settings, the observed clustering of raw message events may reflect the superposition of these distinct temporal processes rather than a single process exhibiting complex temporal dependence.



Formally, we assess whether conversational episodes, rather than raw messages, are more reasonably approximated by the constant-intensity Poisson model introduced in \autoref{sec:poisson_null}. To do so, we introduce a burst-collapse procedure. For each ordered dyad $(i,j)$, with message times $T_{ij} = (t_1, \dots, t_{n_{ij}})$, successive messages are assigned to the same conversational episode whenever their interarrival time $\tau_{ij,k}$ does not exceed a threshold $\delta$, and each episode is represented as a single communication event at the time of its first message. As episodes are defined separately for each direction, a rapid back-and-forth exchange between two actors contributes one episode in each direction. This treats rapid within-conversation exchanges as local conversational microstructure and compresses them into representative interaction episodes, abstracting away fine-scale reply dynamics not central to the network-level communication structure of interest.

\autoref{fig:compression} displays boxplots assessing the effect of the burst-collapse procedure under progressively wider compression windows of 1 minute, 1 hour, 1 day, and 1 week. Specifically, we report the dyad-level Kolmogorov--Smirnov (KS) statistic measuring distributional deviation from the $\mathrm{Exp}(1)$ benchmark, and the retained number of interarrivals per dyad. We plot dyads with at least 10 interarrivals in the raw data and at least three at the corresponding compression level, since fewer observations provide too little information to assess distributional shape.

The KS statistic declines through the first three compression levels, indicating that the interarrival distributions move closer to the exponential benchmark. The largest reductions occur over the initial two compression stages (Raw $\rightarrow$ 1m and 1m $\rightarrow$ 1h), suggesting that much of the observed clustering is attributable to very short-timescale conversational exchanges. Beyond the 1-hour to 1-day range, however, further compression yields diminishing improvements in fit while continuing to reduce the number of retained interarrivals per dyad. Although the median KS statistic decreases slightly from 1 hour to 1 day, it rises marginally again at the 1 week compression level. Given the corresponding reduction in retained observations, this behaviour may partly reflect finite-sample effects.

Overall, burst collapse materially reduces apparent non-Poisson behaviour, though increasingly aggressive compression does so at a corresponding data cost, reducing the effective sample size available for dyadic interarrival analysis. The largest reductions in deviation occur over the initial compression stages, with further compression providing diminishing improvements while continuing to reduce the available observations. This trade-off between a smaller distance from the exponential reference and the number of retained observations provides a simple data-driven criterion for selecting the collapse threshold: the point beyond which further compression yields little additional improvement in fit. For the Conti network this occurs at approximately 1 hour, and unless otherwise stated subsequent analyses use conversational episodes obtained with this threshold as the communication events.

At the selected one-hour threshold, \autoref{fig:interrivals_collapsed} shows the resulting dyadic interarrival distributions alongside the raw-message distributions in \autoref{fig:interrivals_raw}. The excess of short normalised interarrivals is markedly reduced, while the slow tail decay also becomes less pronounced. The collapsed curves therefore lie closer to the Exp(1) interarrival distribution predicted by the Poisson null model, though visible departures remain.

\section{Communication Regimes}
\label{sec:regimes}

\begin{figure*}[h] 
    \centering
    \includegraphics[width=\linewidth]{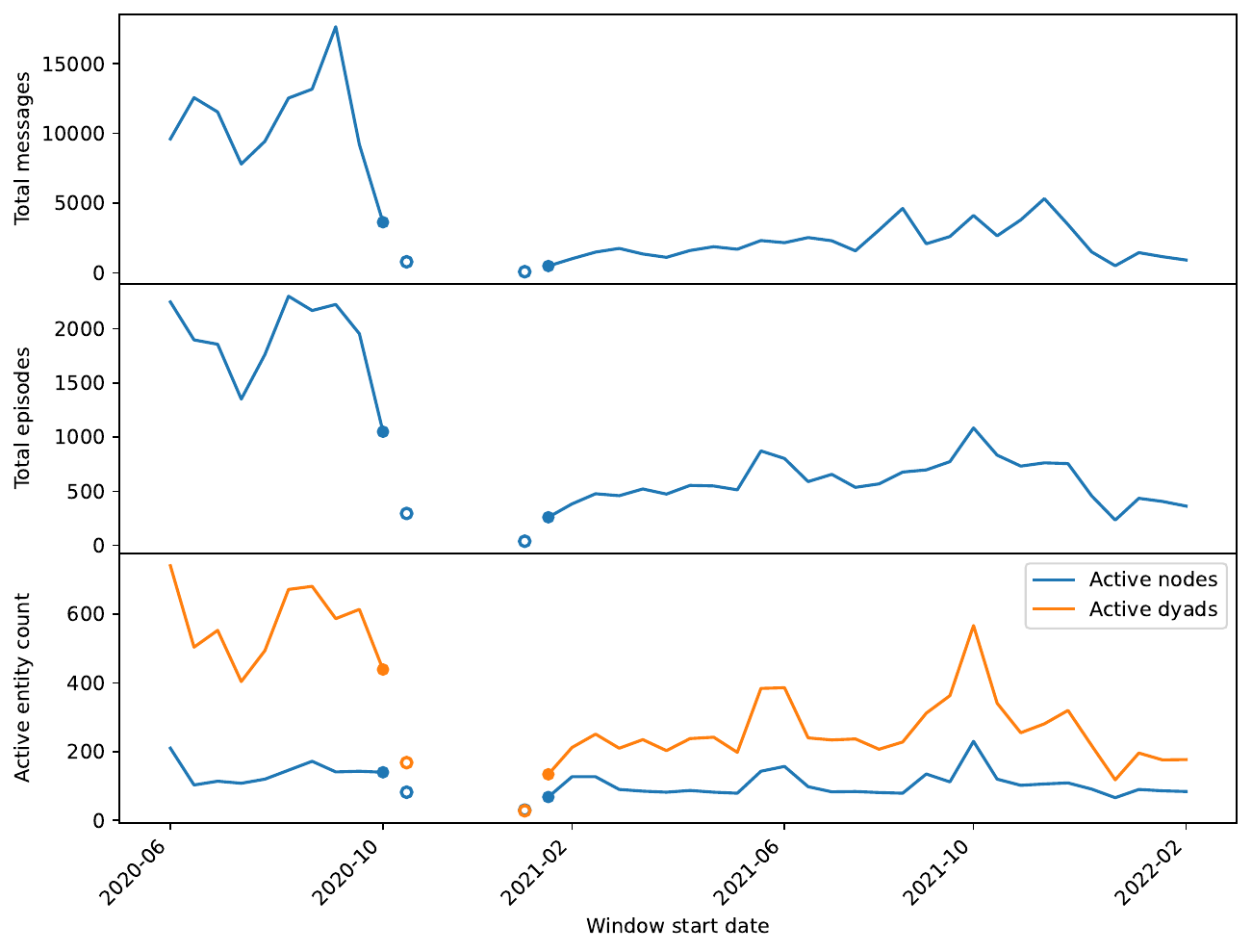} 
    \caption{Temporal activity profile of the communication network across consecutive 14-day windows. Top: raw message counts. Middle: conversational episode counts after 1-hour burst collapse. Bottom: active nodes and active directed dyads. The break denotes the interval without available communication records. Open circles mark adjacent windows that may be partially observed; filled circles mark the ends of the connected series.}
    \label{fig:activity_profile}
\end{figure*}

\begin{figure*}
    \centering
    \includegraphics[width=0.6\linewidth]{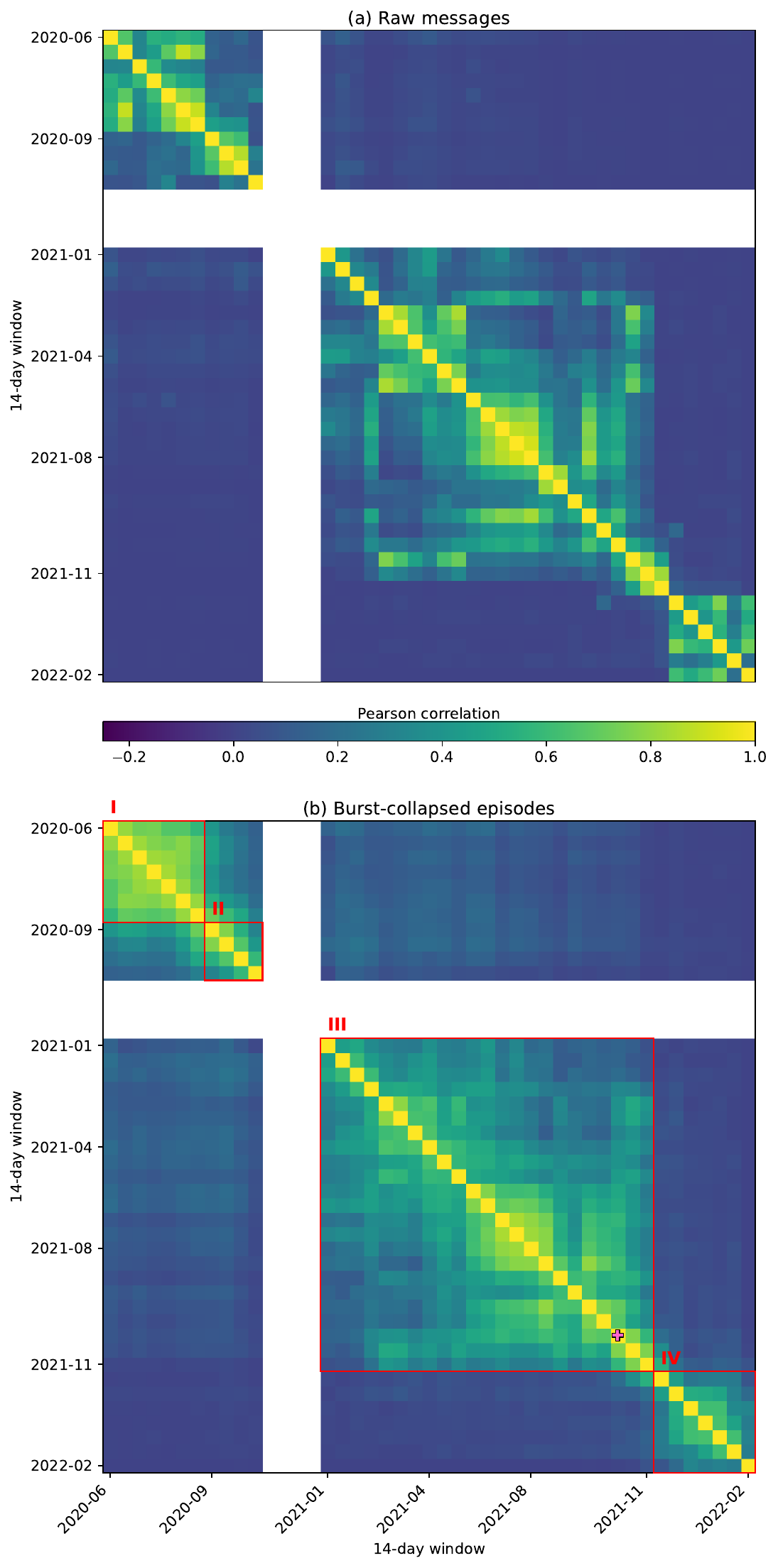}
    \caption{Pairwise Pearson correlations between weighted adjacency matrices formed from consecutive 14-day windows, using raw messages (top) and one-hour conversational episodes (bottom). Brighter colours indicate greater similarity. Red boxes (I--IV) in the bottom panel mark the proposed regimes, and the magenta marker indicates the window containing the initial migration messages discussed in \autoref{sec:regimes}.}
    \label{fig:adj_corr}
\end{figure*}

\subsection{Regime Identification}

The aggregate network representations considered previously are constructed by projecting the timestamped communication events onto a single adjacency matrix over the full observation period. Such representations are often useful for characterising the overall structure of a communication system. However, they also implicitly assume that the large-scale interaction structure remains reasonably stationary through time, such that a single network summary remains meaningful over the observation window.

This assumption of approximate global stationarity may be reasonable for some systems, such as large-scale Internet traffic or online communication platforms. While such systems may exhibit substantial short-timescale variation at the level of individual interactions, the broader structural properties of the network, including overall participation rates, activity volumes, and large-scale connectivity patterns, may remain comparatively stable over longer periods.



In contrast, small to moderate-sized operational or actor-coordination communication networks may exhibit substantial temporal variation in communication intensity, active member sets, and interaction patterns. We begin with the simplest temporal view of the Conti network: its aggregate activity across consecutive 14-day windows. \autoref{fig:activity_profile} shows marked changes in message volume and in the numbers of active nodes and dyads. Recorded activity declines sharply after mid-2020 and later recovers to a lower and more variable level, while active node and dyad counts recover more strongly than message volume. The intervening trough includes a gap in the available records and should not be interpreted entirely as organisational inactivity.

Even this coarse view does not support an approximately constant level of recorded activity across the full period. It does not, however, show whether the pattern of interactions changes. We therefore compare windowed adjacency matrices to examine whether the network is better described as a sequence of locally stable or semi-stationary interaction regimes.

Specifically, we seek to identify a sequence of temporal boundaries

\[
t_1 < t_2 < \dots < t_k
\]
such that the communication structure remains approximately stable within each corresponding interval
\[
[t_1,t_2], [t_2,t_3], \dots, [t_{k-1},t_k],
\]
while differing more substantially between intervals. Equivalently, we seek periods for which the corresponding windowed adjacency matrices may be regarded as approximately locally stationary representations of the underlying communication structure.

Recall that, for any interval $[t_0,t_1]$,
\[
W^{[t_0,t_1]}_{ij}
=
\sum_{(i,j,t)\in E}
\mathds{1}(t_0 \leq t \leq t_1)
\]
records the number of messages sent from actor $i$ to actor $j$ during the corresponding window, with binary projection
\[
A^{[t_0,t_1]}_{ij}
=
\mathds{1}(W^{[t_0,t_1]}_{ij} > 0).
\]


To identify locally stable communication regimes, we partition the observation period into $N$ consecutive fixed-width temporal windows of duration $\Delta t$. This yields a sequence of intervals
\[
[0,\Delta t], [\Delta t,2\Delta t], \dots, [(N-1)\Delta t, N\Delta t].
\]

For each window, we construct the corresponding weighted adjacency matrix
\[
W^{[(n-1)\Delta t,\;n\Delta t]},
\qquad n = 1,\dots,N.
\]
We then compare adjacency structure across time by computing pairwise similarities between flattened weighted adjacency matrices formed from different windows,
\[
\delta A_{k,m} =\mathrm{sim}\!\left(
W^{[(k-1)\Delta t,\;k\Delta t]},
W^{[(m-1)\Delta t,\;m\Delta t]}
\right).\]

Periods exhibiting relatively consistent large-scale interaction structure are therefore expected to produce clusters of comparatively high cross-window similarity, while organisational transitions or operational disruption may produce sharper structural discontinuities between windows.

Under the Poisson null model, weighted adjacency matrices from any two fully observed windows of equal duration $\Delta t$ have the same distribution:
\begin{equation}
    W^{[t_1,t_1+\Delta t]}
    \overset{d}{=}
    W^{[t_2,t_2+\Delta t]}.
\end{equation}
The two matrices are also independent when the windows do not overlap. Pairwise similarities may vary through sampling, but should show no systematic dependence on when the windows occur. Sustained blocks of higher within-period similarity therefore suggest changes in dyadic communication rates.



\autoref{fig:adj_corr} displays similarity matrices for raw messages (top) and one-hour conversational episodes (bottom). Each matrix reports pairwise Pearson correlations between weighted adjacency matrices constructed from consecutive $\Delta t=14$ day windows. Brighter regions correspond to periods exhibiting comparatively similar large-scale communication structure, while darker regions indicate reduced structural similarity between windows.

Several broad blocks of elevated within-period similarity are visible in both panels, suggesting relatively stable communication regimes separated by sharper structural transitions. These blocks are more coherent after burst collapse, with correlations between windows within the proposed regimes more consistently elevated, particularly in the early period and throughout much of 2021. Averaged equally across the four marked regimes, mean within-regime correlation rises from 0.419 to 0.519, while the mean within-regime standard deviation falls from 0.226 to 0.136 (\autoref{tab:regime_summary}). The broad boundaries remain visible in the raw-message panel, suggesting that collapse stabilises the longer-timescale pattern rather than producing a different regime structure. In particular, the network appears to exhibit an early high-activity phase during mid-2020, followed by a shorter transitional period preceding a prolonged gap in the available communication records between late 2020 and early 2021. A larger and comparatively stable interaction regime is then observed throughout much of 2021, before a further transition into a shorter final regime near the end of the observation period. \autoref{tab:regime_summary} quantifies these differences. Daily activity declines from 788.8 messages and 140.8 conversational episodes in regime I to 162.1 messages and 44.2 episodes in regime III, then falls further in regime IV. Regime III nevertheless contains the largest number of distinct observed actors (320), although its longer duration means this count should not be interpreted as a higher daily participation rate.

As a complementary comparison, treating all distinct pairs of windows in the marked regimes as one pooled sample, mean within- and between-regime correlations are 0.328 and 0.026 for raw messages, compared with 0.428 and 0.085 for conversational episodes. Averaging over both types of pair gives global average correlations of 0.137 and 0.211, respectively.


\begin{table*}[h]
\centering
\caption{Summary statistics across the identified communication regimes. Daily rates and active-node counts use the day-level regime cutoffs. Correlations report the mean (SD) of Pearson correlations between distinct pairs of 14-day weighted adjacency matrices within the regime blocks marked in \autoref{fig:adj_corr}.}
\label{tab:regime_summary}
\begin{tabular}{llrrrrrr}
\toprule
Regime & Date range & Days & Msg/day & Episodes/day & Active nodes
       & Raw corr. & Episode corr. \\
\midrule
I   & 21 Jun--18 Sep '20       & 88.4  & 788.8 & 140.8 & 263 & 0.560 (0.182) & 0.744 (0.066) \\
II  & 18 Sep--16 Nov '20       & 59.7  & 640.5 & 111.3 & 172 & 0.455 (0.280) & 0.546 (0.151) \\
III & 29 Jan--1 Dec '21        & 306.0 & 162.1 & 44.2  & 320 & 0.303 (0.212) & 0.403 (0.169) \\
IV  & 1 Dec '21--2 Mar '22     & 91.8  & 121.6 & 32.4  & 148 & 0.359 (0.230) & 0.383 (0.158) \\
\bottomrule
\end{tabular}
\end{table*}


\subsection{Regime Transitions and Platform Migration}

\begin{figure*}[t]
\centering

\begin{tcolorbox}[
colback=gray!5,
colframe=black!60,
title= migration instruction message,
fonttitle=\bfseries]

\small

EVERYONE SHOULD DO THIS, otherwise the toad will stop working for you. the new toad domain version 3 odw5mdwotufuxxrgw3pvqjjuze3e33bylylkl667h4nefwiimwqsumyd.onion we register it in the account settings - connection - server before this on this tab, check the box Manually specify the server address and port standard port 5222 IMPORTANT, everything else remains, the account lags behind with the old domain... only the domain for connection registers!!! When you do, write me back that everything is OK
\end{tcolorbox}

\caption{Migration instruction distributed to Conti members announcing the transition to the version~3 TOAD communication server.}
\label{fig:migration_message}

\end{figure*}

The 14-day adjacency similarity analysis identifies broad transition periods, but does not locate a precise day on which communication structure changed. Even if the underlying structure changed abruptly, its effect on the observed network would become identifiable only as subsequent interactions accumulated. For the subsequent regime-conditioned analyses, we therefore place day-level boundaries within these transition periods using organisational activity observed nearby.

The most apparent transition visible in \autoref{fig:adj_corr} occurs near the boundary between the third and fourth communication regimes during the late stages of the observation period. Inspection of the underlying message activity around this period reveals behaviour indicating an organised platform migration event. 

Specifically, a coordination message, reproduced in \autoref{fig:migration_message}, begins rapidly circulating throughout the network providing instructions for migration to an alternative communication server. The message contains a replacement onion-domain address together with instructions for manually updating the XMPP server connection settings used by the platform. The transmission behaviour associated with this message is itself highly unusual relative to typical communication patterns observed elsewhere in the dataset. Of the 168 observed instances of the message, 148 originate from a single actor, ``defender'', while the remaining transmissions also involve ``defender'' as the recipient. Furthermore, 139 of the observed transmissions occur within a four-minute interval, with the remaining messages distributed more sparsely across later periods. This behaviour suggests an unusually rapid and highly centralised coordination event. Taken together, these observations indicate that the regime III--IV transition corresponds to a broader organisational migration, in which communication activity progressively shifted away from the original platform captured in the leaked dataset.

Notably, the initial circulation burst of the migration message occurs on 22 October 2021, marked in \autoref{fig:adj_corr}. While this occurs relatively close to the observed regime III--IV transition period, it does not align exactly with the primary structural breakpoint visible in the adjacency similarity analysis, which occurs after the initial migration-message circulation event. The delay is consistent with a gradual change in platform use. Although many actors sharply reduce their activity after receiving the migration instructions, this response is not uniform: some highly active actors continue communicating on the original platform for substantial periods. Their continued activity can preserve the earlier weighted adjacency structure even as other participants move away. Some coordination may also have shifted to the replacement infrastructure, which is not observed in the leaked dataset. The structural transition may therefore reflect the later consolidation of the replacement platform as the primary coordination channel, rather than the initial dissemination of the migration instructions.

To place a day-level boundary for the regime-conditioned analyses, we examine the underlying messages around the regime III--IV transition. On 29--30 November 2021, a flurry of encrypted messages involved ``defender'', the principal disseminator of the October migration instructions. Their content cannot be assessed, but the timing is consistent with a later stage of the migration process. We place the regime III--IV cutoff on 1 December, immediately after this activity. The earlier regime I--II boundary is placed in the same manner. On 17 September 2020, a request for project-management and programming-team experience was sent to 64 recipients; we place the cutoff on 18 September. In both cases, the day-level boundary is a working partition within a broader transition identified from the 14-day similarity analysis, rather than a precisely timed change in network structure.

\section{Participation Dynamics and Observability}
\label{sec:accum}
\subsection{Participant Discovery}

\begin{figure*}[!t]
\centering


\begin{subfigure}{0.48\textwidth}
    \centering
    \includegraphics[width=\linewidth]{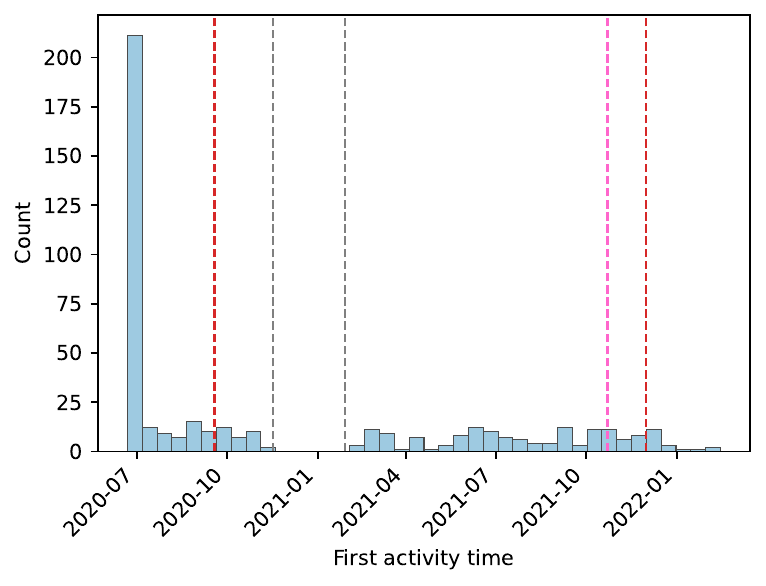}
    \caption{Node first activity time.}
\end{subfigure}
\hfill
\begin{subfigure}{0.48\textwidth}
    \centering
    \includegraphics[width=\linewidth]{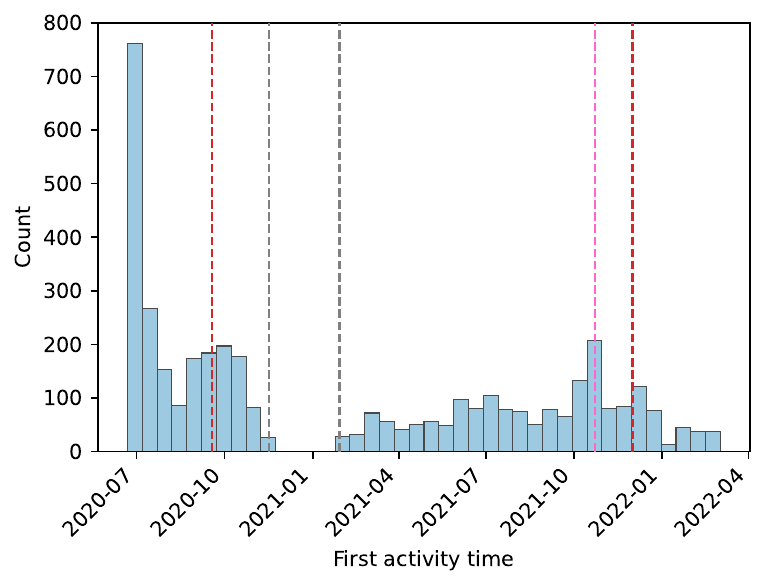}
    \caption{Dyad first activity time.}
\end{subfigure}

\vspace{0.5em}


\begin{subfigure}{0.48\textwidth}
    \centering
    \includegraphics[width=\linewidth]{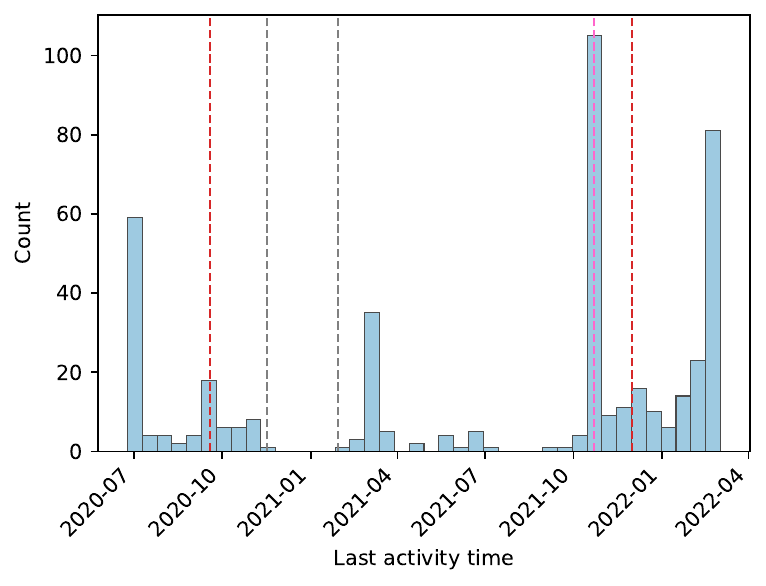}
    \caption{Node last activity time.}
\end{subfigure}
\hfill
\begin{subfigure}{0.48\textwidth}
    \centering
    \includegraphics[width=\linewidth]{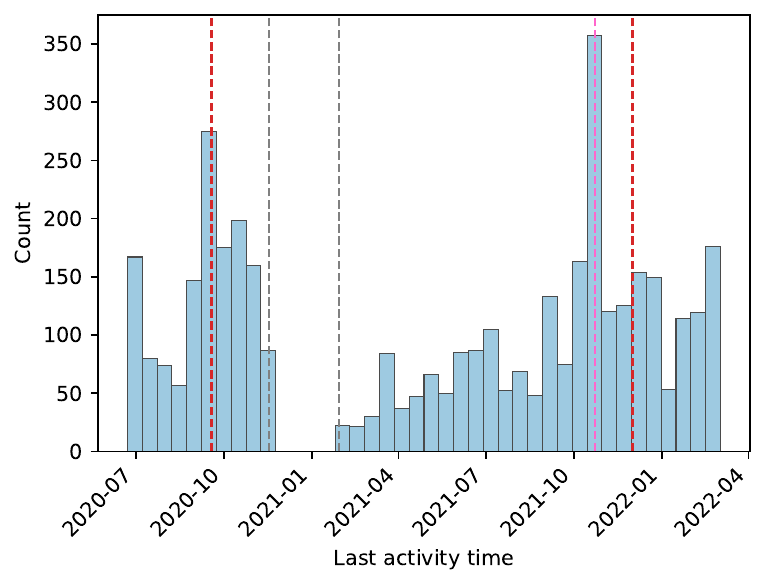}
    \caption{Dyad last activity time.}
\end{subfigure}

\vspace{0.5em}


\begin{subfigure}{0.48\textwidth}
    \centering
    \includegraphics[width=\linewidth]{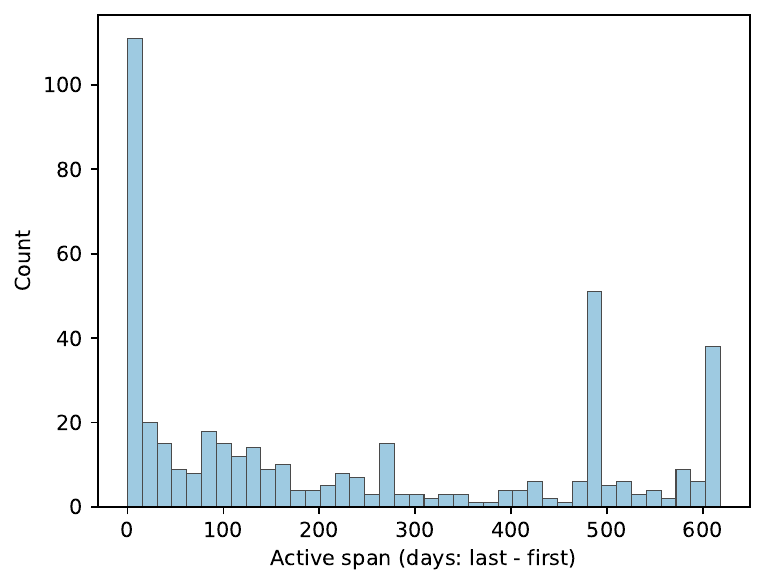}
    \caption{Node active span.}
\end{subfigure}
\hfill
\begin{subfigure}{0.48\textwidth}
    \centering
    \includegraphics[width=\linewidth]{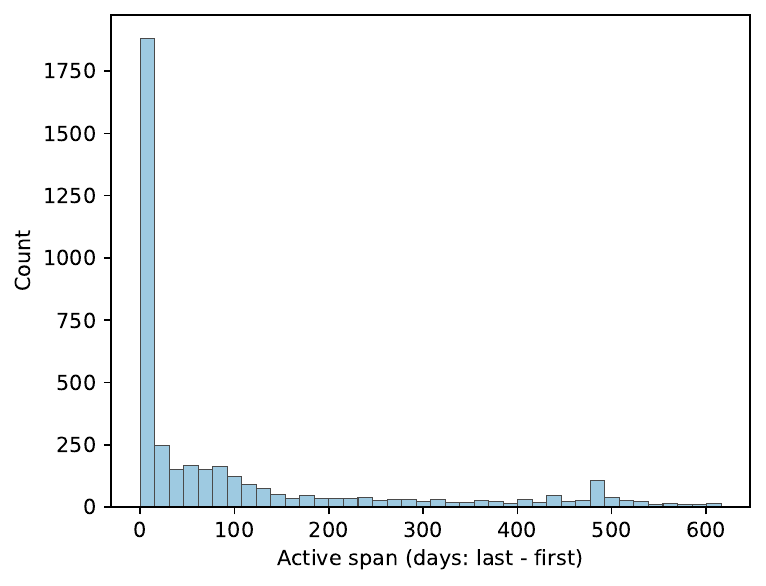}
    \caption{Dyad active span.}
\end{subfigure}

\caption{Distributions of first activity time, last activity time, and active span for nodes and dyads. First and last activity are shown in calendar time; active span is the elapsed time between first and last observed activity. In the calendar-time panels, red dashed lines mark the selected regime boundaries, the magenta dashed line marks the initial migration-message circulation on 22 October 2021, and grey dashed lines bound the gap in the available records.}

\label{fig:lifespan_activity}
\end{figure*}

The preceding sections identified substantial variation in communication structure across multiple temporal scales. At the coarsest level, the adjacency similarity analysis revealed distinct communication regimes separated by periods of structural reorganisation. A natural subsequent question is whether communication behaviour appears approximately stationary within these regimes, or whether meaningful population turnover continues to occur even after conditioning on the broader regime structure.

This question is closely related to the problem of network observability. In practice, analysts rarely observe a complete communication network directly. Instead, network structure is inferred from a finite collection of observed messages accumulated over some period of time. The extent to which additional observation effort continues to reveal previously unseen actors therefore provides information about both the local stationarity of the communication process and the completeness of the recovered network representation.

Species accumulation methods \cite{good1953population,chao1984nonparametric,colwell1994estimating} provide a useful framework for examining this problem. Rather than focusing on changes in communication topology, accumulation analysis considers the rate at which previously unobserved actors continue to appear as additional communication activity is sampled. Under a comparatively stationary communication process, the rate of new actor discovery would be expected to decline as observation effort increases. Conversely, persistent discovery of new actors may indicate ongoing population turnover, temporal heterogeneity of dyad activity, or incomplete observation of the underlying communication population.



\begin{figure*}[t]
\centering

\begin{subfigure}{0.49\linewidth}
    \centering
    \includegraphics[width=\linewidth]{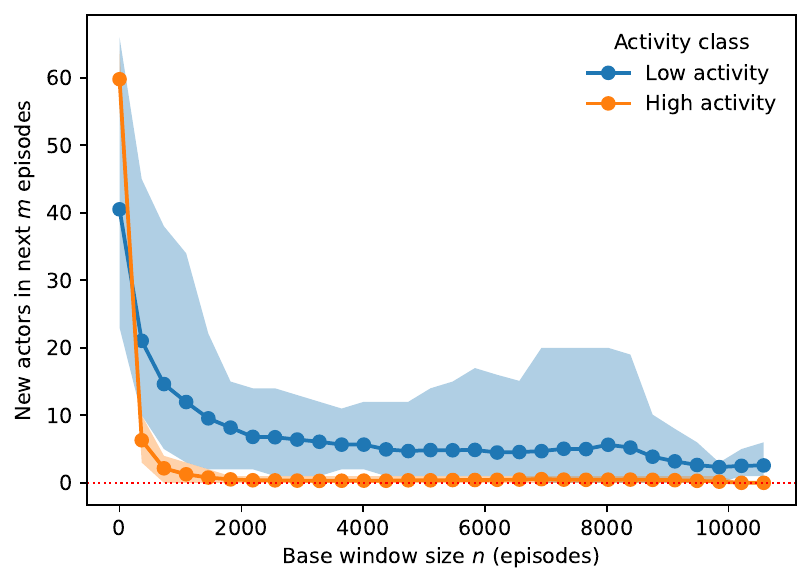}
    \caption{Regime I}
\end{subfigure}
\hfill
\begin{subfigure}{0.49\linewidth}
    \centering
    \includegraphics[width=\linewidth]{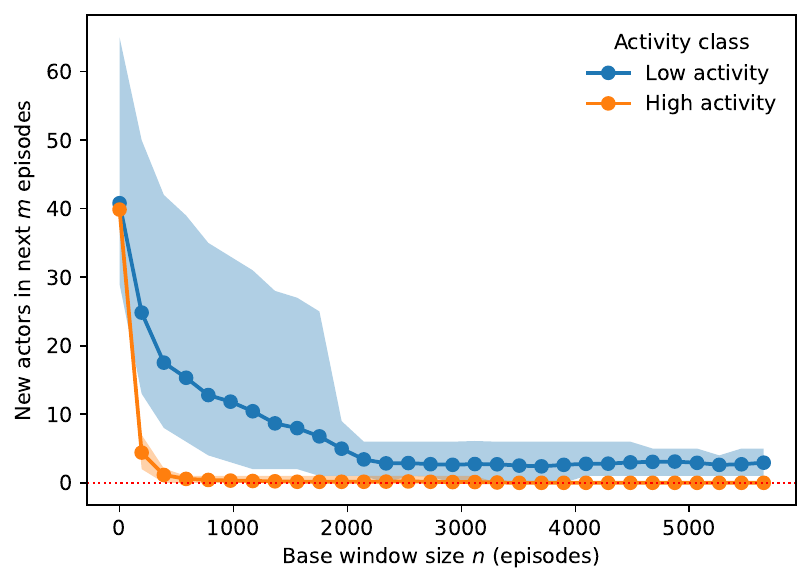}
    \caption{Regime II}
\end{subfigure}

\vspace{0.5em}

\begin{subfigure}{0.49\linewidth}
    \centering
    \includegraphics[width=\linewidth]{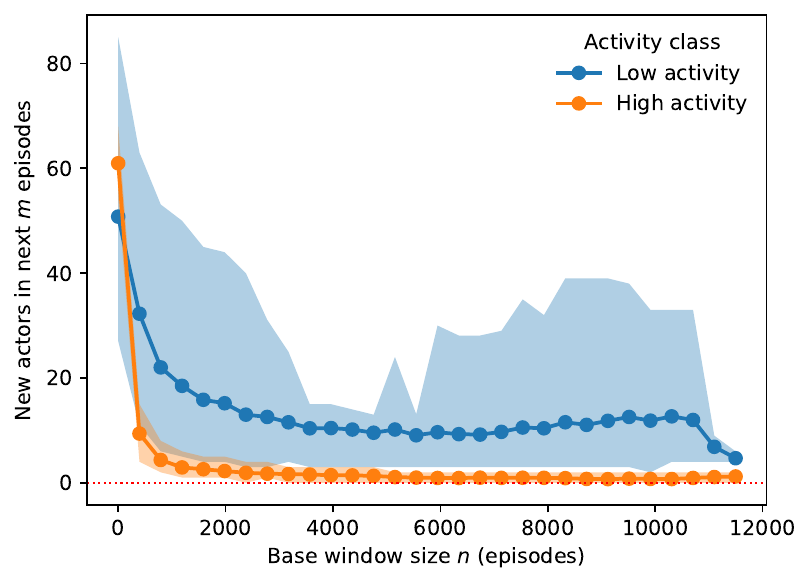}
    \caption{Regime III}
\end{subfigure}
\hfill
\begin{subfigure}{0.49\linewidth}
    \centering
    \includegraphics[width=\linewidth]{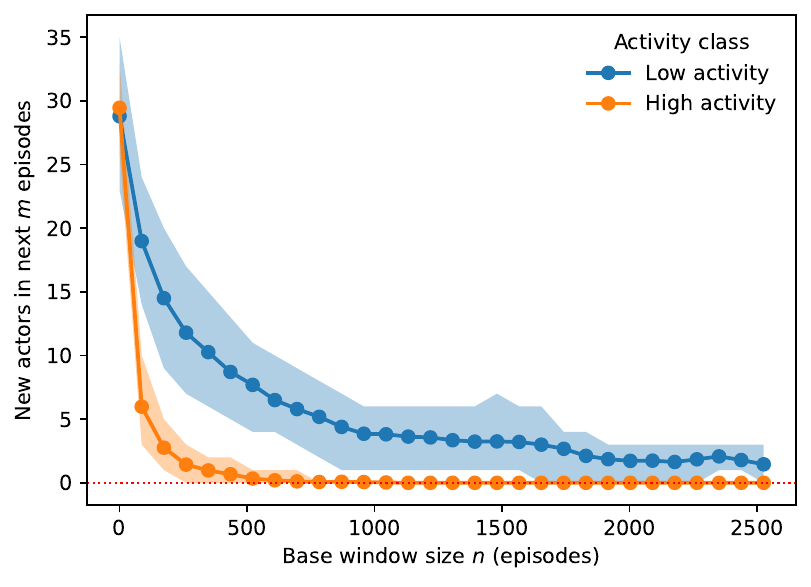}
    \caption{Regime IV}
\end{subfigure}

\caption{Look-forward actor discovery across the four communication regimes. Each curve shows the mean number of previously unseen actors appearing in the next $m$ conversational episodes after a base window of $n$ episodes, averaged over sampled starting positions. The horizon $m$ is 5\% of the episodes in each regime. High- and low-activity classes are defined separately within each regime, with the top 25\% of actors by episode participation assigned to the high-activity class. Shading shows the 10th--90th percentiles across sampled starting positions.}

\label{fig:add_m_regimes}

\end{figure*}

\autoref{fig:add_m_regimes} presents look-forward actor discovery curves for each communication regime. Within a regime, let $e_1,\ldots,e_N$ be the conversational episodes in temporal order, and let $V(k,n)$ denote the actors observed in episodes $e_{k+1},\ldots,e_{k+n}$. The number of additional actors revealed by a further $m$ episodes is:
\begin{equation*}
    \left|V(k,n+m)\setminus V(k,n)\right|
    =\left|V(k,n+m)\right|-\left|V(k,n)\right|.
    \label{eq:forward_discovery}
\end{equation*}
We fix $m$ at 5\% of the episodes in each regime. For each base window
size $n$, we sample starting positions $k$ uniformly from those for
which all $n+m$ episodes are observed. The curves show the mean number
of additional actors across starting positions, with shading indicating the 10th--90th percentiles. Within each regime, actors are ranked by their total number of appearances as sender or recipient in conversational episodes. The top 25\% of actors active in that regime form the high-activity class; the remaining actors form the low-activity class.


Across all four regimes, highly active participants become observable rapidly, with the discovery curves reaching saturation after only a modest fraction of each regime. The comparatively narrow percentile ranges further indicate that these actors are identified consistently regardless of the initial observation point, suggesting that the principal communication structure is robustly observable throughout each regime. In contrast, less active participants exhibit a substantially slower decay in discovery rate together with considerably greater variability between observation windows. While part of this slower decay is expected from their lower communication frequency, the extended non-zero tail indicates sustained discovery of previously unseen actors throughout each regime, rather than a simple reduction in observation rate. This interpretation is further supported by the broad distribution of first observed activity times shown in \autoref{fig:lifespan_activity}, indicating that many lower-activity participants first become active well after the commencement of each communication regime.

\subsection{Discovery Under a Stationary Null}

\begin{figure*}[t]
\centering

\begin{subfigure}{0.49\linewidth}
    \centering
    \includegraphics[width=\linewidth]{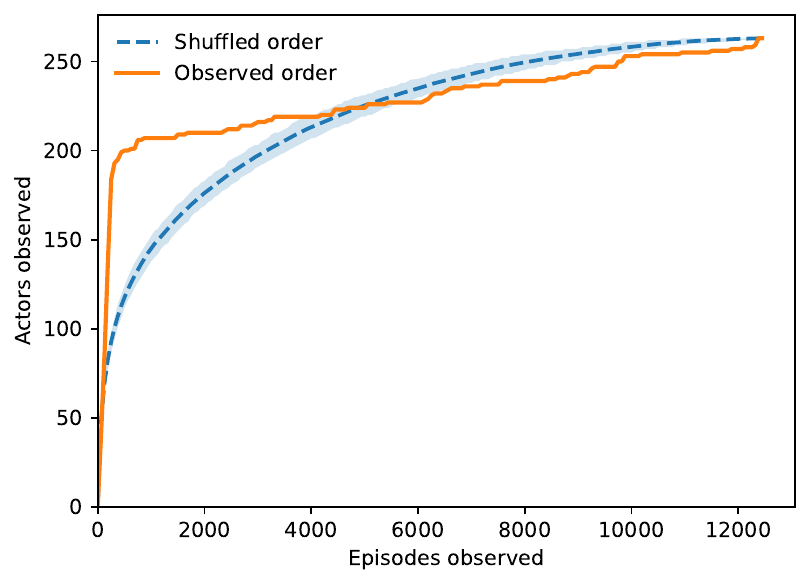}
    \caption{Regime I}
\end{subfigure}
\hfill
\begin{subfigure}{0.49\linewidth}
    \centering
    \includegraphics[width=\linewidth]{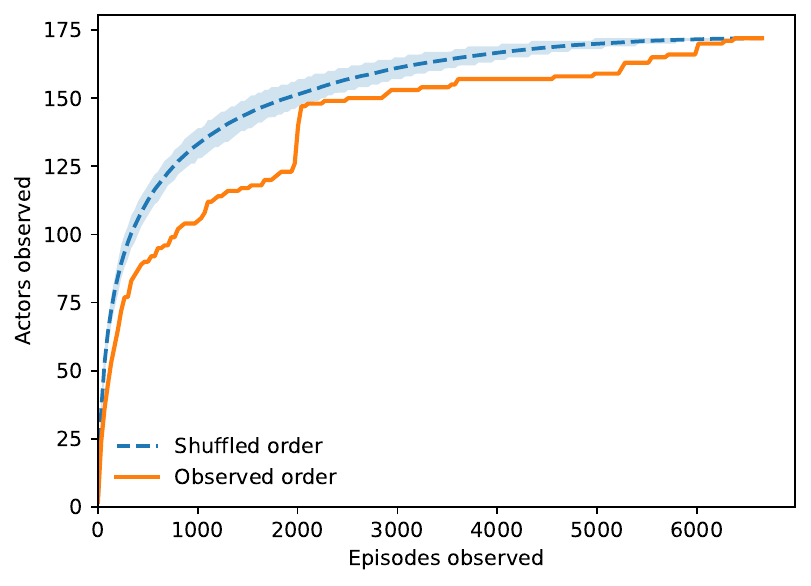}
    \caption{Regime II}
\end{subfigure}

\vspace{0.5em}

\begin{subfigure}{0.49\linewidth}
    \centering
    \includegraphics[width=\linewidth]{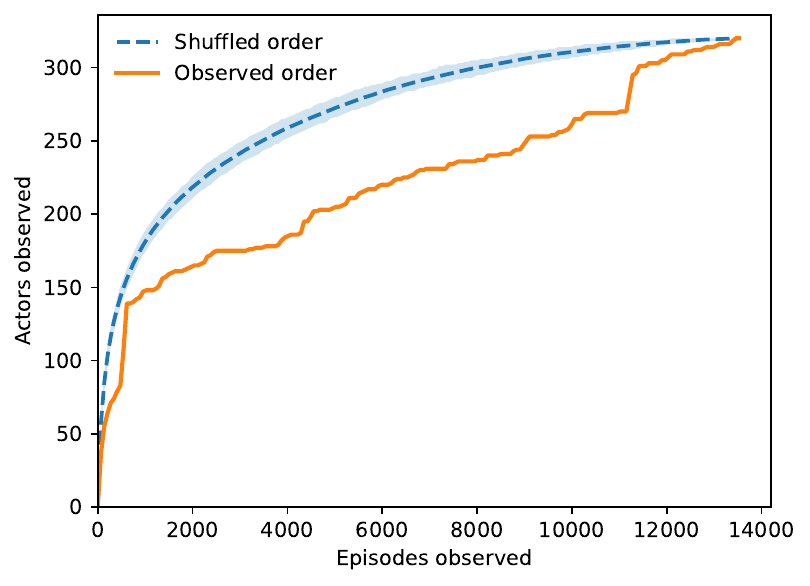}
    \caption{Regime III}
\end{subfigure}
\hfill
\begin{subfigure}{0.49\linewidth}
    \centering
    \includegraphics[width=\linewidth]{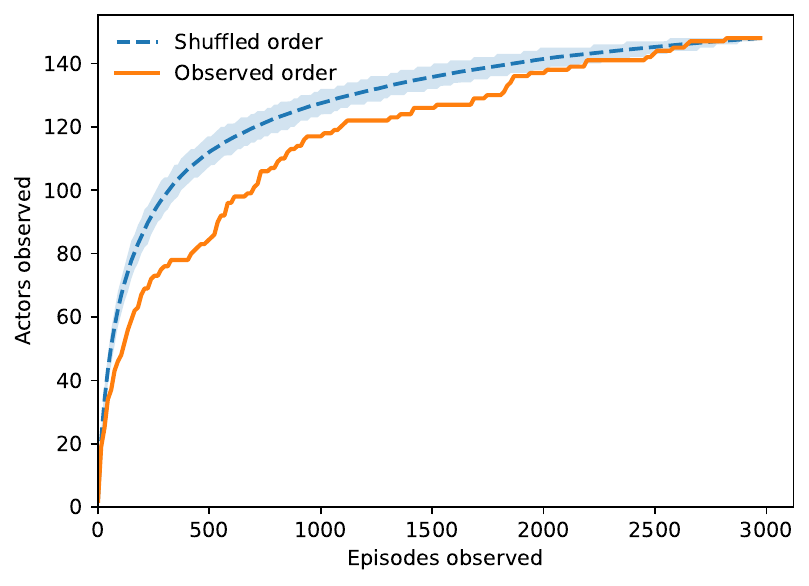}
    \caption{Regime IV}
\end{subfigure}

\caption{Node accumulation curves across the identified communication regimes. The temporal curve shows cumulative node discovery in observed episode order, while the shuffled baseline randomises episode order while preserving the sender and recipient of each episode.}

\label{fig:node_accum_temporal_shuffled}

\end{figure*}

\autoref{fig:node_accum_temporal_shuffled} compares cumulative participant discovery in the observed episode order with a shuffled baseline. Within each regime, the baseline randomly reorders the episodes while preserving the sender and recipient. It therefore preserves the aggregate communication network within each regime, while removing the observed temporal ordering of interactions. This baseline corresponds to the Poisson null model of \autoref{sec:poisson_null}, with dyad-specific rates estimated separately within each regime. Conditional on the observed dyad counts, the event times of a homogeneous Poisson process are independent and uniformly distributed, so every ordering of the observed communication events is equally likely.
Across the identified communication regimes, both the empirical and shuffled curves exhibit an initial period of rapid accumulation corresponding to the early discovery of the principal communicators. Under the shuffled baseline, this transitions smoothly towards saturation as progressively fewer undiscovered participants remain. In contrast, the empirical curves typically exhibit a markedly slower decay following the initial accumulation period, with extended intervals of near-linear growth interspersed with comparatively rapid bursts of participant discovery.

While some reduction in accumulation rate is expected as the remaining undiscovered participants communicate less frequently, the sustained near-linear growth observed in the empirical curves is inconsistent with a homogeneous communication process. Rather, participant discovery remains concentrated within particular periods of each communication regime, indicating that the temporal ordering of communication events contributes substantially to participant observability beyond differences in aggregate communication activity alone.

\section{discussion}

\subsection{Accounting for Temporal Heterogeneity}

The analyses presented in this work identify several well-characterised forms of temporal heterogeneity that can be explicitly incorporated into the analysis of cybercriminal communication networks. These include short-timescale conversational structure and longer periods of comparatively stable network organisation.



At the shortest temporal scales, the burst-collapse results indicate that much of the observed temporal heterogeneity can be accounted for through a straightforward aggregation of temporally adjacent messages into conversational episodes, with the collapse threshold selected as the point beyond which agreement with the Poisson null model no longer materially improves. Conceptually, while messages constitute the directly observed communication events, organisational coordination is more naturally expressed through conversations than the individual messages and replies that comprise them. The organisational event of interest is therefore often the occurrence of communication between actors rather than the number of messages exchanged once contact has been established. Taken together, the marked reduction in short-horizon temporal heterogeneity following burst collapse, together with the clearer isolation of higher-level organisational dynamics in subsequent analyses, corroborates the qualitative intuition that conversational episodes constitute the natural event representation of organisational communication.

At longer temporal scales (weeks to months), the identified communication regimes provide periods over which the dynamic communication process can be more reasonably approximated as a single network. Although temporal variation remains within each regime, the observed adjacency structure is substantially more stable within regimes than between them. Consequently, much of the longer-timescale heterogeneity evident under a single temporally aggregated network can be organised into periods of comparatively stable communication structure through regime partitioning. These periods therefore provide natural observational units over which communication networks can be analysed. In contrast, predefined observation periods or fixed-width temporal windows are unlikely to align with the underlying phases of organisational communication, potentially partitioning coherent regimes or combining distinct organisational states within a single network representation \cite{krings2012effects}. For longitudinal intelligence collection and network monitoring, observation periods should therefore be selected to align as closely as possible with underlying communication regimes rather than arbitrary calendar intervals \cite{holme2012temporal}. The principal challenge, however, lies not in analysing communication once such periods have been identified, but in determining when they begin and end.

\subsection{Residual Temporal Heterogeneity}
While communication regimes provide useful observational units once identified, determining their timing as they occur remains considerably more challenging. Unlike conversational episodes, which arise from well-defined temporal proximity, regime transitions reflect organisational processes that need not manifest as a single discrete event within the observed network. The dissemination of migration instructions associated with the regime III--IV transition represents an unusually strong qualitative indicator of organisational change, yet even in this case the event does not correspond exactly to the identified regime boundary. This suggests that the observed regime boundary need not coincide with the initiating organisational event, but instead reflects the point at which the effects of that change become apparent within the observed communication network. Retrospective identification is, however, substantially more tractable than prospective detection. 
We present a method in which regime boundaries emerge clearly from the evolving adjacency structure itself, allowing coherent observational periods to be identified directly from the data rather than requiring detailed qualitative interpretation of individual events. Such ex post identification is therefore well suited to the analysis of historical communication datasets. In contrast, operational monitoring requires regime transitions to be inferred as they unfold, before the resulting communication structure is fully observed, making prospective regime detection substantially more challenging.

Even after accounting for conversational microstructure and communication regimes, meaningful temporal heterogeneity remains within the communication process. The actor discovery analyses indicate that this residual variation is not distributed uniformly across the network. Highly active participants are rapidly and consistently observable throughout each regime, whereas sustained actor discovery is driven primarily by lower-activity participants appearing during comparatively brief periods of communication. Consequently, the remaining uncertainty within communication regimes appears to be concentrated within the network periphery, while the organisational core remains comparatively well observed.

These findings have practical implications for intelligence collection and longitudinal network monitoring. The rapid and persistent observability of highly active participants suggests that the most active elements of an organisation may often be characterised relatively early within an observation period. In contrast, sustained participant discovery is concentrated within the network periphery, indicating that incomplete observation is likely to disproportionately affect understanding of lower-activity participants even when the principal communication structure has been successfully characterised.

\section{Conclusion}

This study investigated temporal heterogeneity within the Conti ransomware communication network through analyses spanning conversational microstructure, network reconfiguration, and participant dynamics. 
Collectively, these analyses show that the temporal homogeneity implicitly assumed by aggregate network analysis is inconsistent with the observed communication at each resolution examined, so that routine choices of event definition and observation period materially shape the network structure that is inferred.

Several of these sources of temporal variation can be explicitly incorporated into network analysis. 
Conversational episodes provide a more natural event representation than individual messages, bringing communication timing closer to a constant-intensity reference and yielding clearer, more stable structure at longer timescales. Communication regimes in turn provide coherent observational periods, with network structure substantially more consistent within regimes than between them. Even within these regimes, however, participant observability remains temporally heterogeneous, with highly active participants becoming rapidly observable while lower-activity participants continue to appear throughout the observation period.
    
Together, these findings suggest that cybercriminal communication networks are more appropriately viewed as dynamically reconfiguring systems than single static organisational structures. More broadly, they demonstrate the importance of temporally resolved network analysis for understanding organisational communication, providing empirical guidance for future temporal network modelling, longitudinal network analysis, and intelligence collection.

\bibliographystyle{IEEEtran}

\bibliography{ref}

\end{document}